\documentclass[12pt]{article}
\usepackage{epsfig}  
\usepackage{amsmath}
\usepackage{amssymb}

\begin{document}
\noindent
\begin{center}
{\Large {\bf Constraints on Interacting Early Dark Energy from a Modified Temperature–Redshift Relation and CMB Acoustic Scales   }}\\
\vspace{2cm}
 ${\bf Yousef~Bisabr}$\footnote{e-mail:~y-bisabr@sru.ac.ir.}\\
\vspace{.5cm} {\small{Department of Physics, Shahid Rajaee Teacher
Training University,
Lavizan, Tehran 16788, Iran}}\\
\end{center}
\begin{abstract}
The Hubble tension, reflecting a persistent discrepancy between early- and late-time determinations of the Hubble constant, continues to motivate extensions of the standard cosmological model that modify pre-recombination physics. In this work, we investigate the cosmological implications of an early dark energy (EDE) scalar field exponentially coupled to the radiation sector prior to recombination. This interaction induces a homogeneous energy exchange that alters the radiation dilution law and leads to a non-standard cosmic microwave background (CMB) temperature–redshift relation of the form $T(z)=T_0(1+z)^{1-\beta}$, where the parameter $\beta$ quantifies the strength of the coupling.

We first derive the modified temperature–redshift relation from the background evolution equations and interpret it in terms of effective photon creation or annihilation. We then study linear scalar perturbations in the tight-coupling regime relevant for CMB acoustic physics. Although the EDE–photon interaction modifies the background radiation density and thermal history, we show that it does not introduce new dynamical degrees of freedom at the perturbation level. Consequently, the structure of the photon–baryon perturbation equations remains unchanged, while key background-dependent quantities, including the sound speed and the comoving sound horizon at recombination, acquire a dependence on the coupling parameter.

We demonstrate that the dominant observational effect of the EDE–photon interaction arises through a shift in the sound horizon, which in turn modifies the angular acoustic scale measured in the CMB anisotropy spectrum. Using the high-precision \textit{Planck} constraint on the acoustic scale, we derive a semi-analytic consistency bound on the coupling strength and find that deviations from the standard temperature–redshift relation are constrained at the level $|\beta|\lesssim10^{-3}$. Our results show that even small departures from adiabatic photon dilution can leave observable imprints on early-Universe probes, providing a well-motivated target for current and future high-precision CMB experiments.
\end{abstract}


~~~~~~~~~~~~~~~~~~~~~~~~~~~~~~~~~~~~~~~~~~~~~~~~~~~~~~~~~~~~~~~~~~~~~~~~~~~~~~~~~~~~~~~~~~~~~~~~~~~~~
\section{Introduction}
The CMB is a thermal radiation from the recombination era at a redshift
$z\approx 1100$ serv serves as a powerful probe of the early Universe. In the standard $\Lambda$CDM model, the CMB temperature is given by the adiabatic relation $T(z)=T_0(1+z)$ with $T_0\approx 2.725~K$
 being the present-day value and reflects a Universe that evolves through radiation, matter and dark energy domination \cite{peeb}. The CMB temperature anisotropies, described by the power spectrum $C_l$, encode information about primordial perturbations, acoustic oscillations in the photon–baryon fluid and the geometry of spacetime and is therefore a powerful probe of cosmological parameters \cite{hu}\cite{planck}. Any deviation from this standard picture either in the temperature-redshift relation or the anisotropies spectrum, signals new physics beyond $\Lambda$CDM which motivates  studies of alternative cosmological evolution.\\
 Scalar fields have long been used to explain phenomena ranging from inflation to dark energy. In the radiation-dominated era, they are typically subdominant and their energy density is rapidly diluted by expansion of the Universe. The coupling of scalar fields with radiation affects this dilution, the background evolution and the perturbation dynamics in the early Universe. Initial applications of such interactions emerge in inflationary reheating where scalar fields like the inflaton decayed into radiation to trigger the hot Big Bang era \cite{kof}. Subsequent research explored interactions between scalar fields and photons or other Standard Model fields (often inspired by string theory or axion-like fields) impacting cosmological observables such as the CMB \cite{carroll}\cite{kap}. Recently, scalar-photon interactions have been invoked to resolve anomalies such as the Hubble tension by leveraging energy exchange between dark sectors and radiation. \cite{sch}\cite{bis0}.\\
 In our previous work \cite{bis1}, we demonstrated that a scalar field endowed with a potential which exponentially coupled to radiation during the radiation-dominated era alters the expansion rate of the Universe. By considering the scalar field as an early dark energy (EDE) field, we showed that the EDE-photon coupling can generate an effective fluid whose equation of state mimics a cosmological constant. Based on this framework, the current study investigates the consequences of such an EDE-photon coupling for two key CMB observables: the temperature-redshift relation and the anisotropies. In standard adiabatic expansion, a non-interacting radiation fluid yields $T(z)\propto (1+z)$.
 The introduction of EDE-photon coupling leads to energy exchange with the possibility that $T(z)$
 differs from the usual expression. Similar deviations, studied in the context of decaying dark energy or photon dissipation context \cite{lima2}\cite{oph} give tight constraints from CMB spectral distortions and moderate redshift temperature observations \cite{fix}\cite{av}. Moreover, the coupling influences the evolution of density perturbations and distorts the CMB anisotropy power spectrum $C_l$ by shifting acoustic oscillations \cite{hart}\cite{gali}.\\
The motivation for this study is twofold: First, exploring how the EDE-photon coupling affects the standard CMB temperature-redshift relation. Second, studying how such a modification in the thermal history and expansion rate can leave detectable signatures in a variety of cosmological observables including shifts in the CMB acoustic features and distortions in the CMB energy spectrum. Based on our earlier work on background evolution, we aim to quantify these effects and examine how current and future observational data can be used to constrain the strength of EDE-photon interactions. This approach provides testable predictions for non-standard cosmological scenarios motivated by the Hubble tension.\\
This paper is organized as follows. In Section 2, we introduce the EDE–photon interacting model and derive the resulting modification to the CMB temperature–redshift relation and show explicitly how deviations from the standard adiabatic scaling are controlled by the rate of energy exchange between the EDE and radiation. In Section 3, we study linear scalar perturbations in the tight–coupling regime prior to recombination. We demonstrate that although the EDE–photon coupling modifies the background radiation density and thermal history it does not introduce new dynamical degrees of freedom at the perturbation level. As a consequence, the structure of the photon–baryon perturbation equations remains unchanged while key physical quantities such as the sound speed and comoving sound horizon acquire a dependence on the coupling parameter. We show how these modifications feed into the acoustic scale that governs the CMB anisotropy spectrum and derive semi–analytic constraints on the coupling strength from the precise Planck measurement of the angular acoustic scale. In Section 4, we summarize our conclusions.

~~~~~~~~~~~~~~~~~~~~~~~~~~~~~~~~~~~~~~~~~~~~~~~~~~~~~~~~~~~~~~~~~~~~~~~~~~~~~~~~~~~~~~~~~~~~~~~~~~~~~~~~~~~~~~~~~~~~~~~~~~~~~~~~
\section{Modifying the CMB temperature-redshift relation}
The action for the system under consideration is given by
\begin{equation}
S= \int d^{4}x \sqrt{-g} \{\frac{1}{2}R
-\frac{1}{2}g^{\mu\nu}\nabla_{\mu}\phi
\nabla_{\nu}\phi-V(\phi)+e^{-\sigma\phi}L_{m}\}
\label{a1}\end{equation}
which contains a minimally coupled scalar field $\phi$ with a potential $V(\phi)$. The scalar field couples with the matter sector via the coupling function $e^{-\sigma\phi}$ with $\sigma$ being a coupling parameter. The Lagrangian density $L_m$ contains the dominant radiation, baryons and cold dark matter (CDM) which are subdominant. We use this framework to model an EDE-photon interaction at early times during the radiation-dominated era before recombination. \\
Due to the EDE-photon coupling, the energy density of radiation $\rho_{\gamma}$ satisfies
\begin{equation}
\dot{\rho}_{\gamma}+4H\rho_{\gamma}=\frac{4}{3}\sigma \dot{\phi}\rho_{\gamma} \label{a3}\end{equation}
Here $H$ is the Hubble parameter which satisfies the Friedmann equation
\begin{equation}
3H^2= e^{-\sigma\phi}\rho_{\gamma}+\rho_{\phi}
\label{a2-1}\end{equation}
The equation of $\phi$ is
\begin{equation}
\ddot{\phi}+3H\dot{\phi}+V(\phi)=-\frac{1}{3}\sigma e^{-\sigma\phi}\rho_{\gamma}
\label{}\end{equation}
The equation (2) can be immediately integrated which yields \cite{bis1}
\begin{equation}
\rho_{\gamma}\propto a^{-4+\epsilon}
\label{}\end{equation}
with
\begin{equation}
\epsilon\equiv \frac{4\sigma\phi}{3\ln a}
\label{a3}\end{equation}
The coupling parameter $\epsilon$ determines the strength and direction of energy transfer between EDE and radiation.  For $\epsilon>0$, energy flows from EDE into the radiation field leading to production of radiation. In this case, the energy density $\rho_{\gamma}$ decreases more slowly compared to the standard adiabatic evolution $\rho_{\gamma}\propto a^{-4}$. Conversely, when $\epsilon<0$ energy transfers out of radiation leading to $\rho_{\gamma}$ to dilute faster. \\
This EDE-photon coupling will generally distort the behaviour of the radiation fluid and in particular the photon temperature-redshift relation away from
its standard evolution. To relate such a deviation to $\epsilon$, we combine $\rho_{\gamma}\propto T^{4}$\footnote{We assume that the photon bath remains in local thermal equilibrium with a vanishing chemical potential \cite{lima1} so that the radiation energy density retains its blackbody form $\rho_{\gamma}\propto T^4$ even in the presence of homogeneous energy exchange.} with (2) which leads to
\begin{equation}
\frac{\dot{T}}{T}+H=\frac{\sigma}{3}\dot{\phi}
\label{a4}\end{equation}
This differential equation governs the rate of change of the CMB temperature $T(z)$. In the limiting case $\sigma=0$, the scalar field does not interact with radiation and (7) gives the standard adiabatic evolution law $T(z)=T_0 (1+z)$ with $T_0$ being an integration constant. In contrast, when $\sigma\neq 0$, the scalar field is coupled to radiation resulting in an exchange of energy between the two components and consequently a deviation from the standard temperature–redshift law. In this case, (7) leads to
 \begin{equation}
 T(z)=T_0 (1+z)f(z)
\label{a5}\end{equation}
where $f(z)$ is a modification function. This equation demonstrates that the CMB temperature-redshift law is modified due to the EDE-photon coupling and the modification is encoded in the function $f(z)$ which is given by
\begin{equation}
 \frac{df(z)}{f(z)}=\frac{\sigma}{3} d\phi
\label{a6}\end{equation}
or, equivalently,
\begin{equation}
f(z)=e^{\frac{\sigma}{3}\int d\phi}
\label{a7}\end{equation}
Using (\ref{a3}) the integrand in (10) can be written as $d\phi\propto d(\epsilon\ln a)$ making explicit that $\epsilon$ is an evolving quantity. In this generic case, the resulting form of $f(z)$ is model-dependent. However, if $\epsilon$ is taken to be constant, the integral simplifies and $f(z)$ assumes the simple power-law form given by
\begin{equation}
f(z)\propto (1+z)^{-\beta}
\label{a7}\end{equation}
\begin{figure}[h]
\begin{center}
\includegraphics[width=0.8\linewidth]{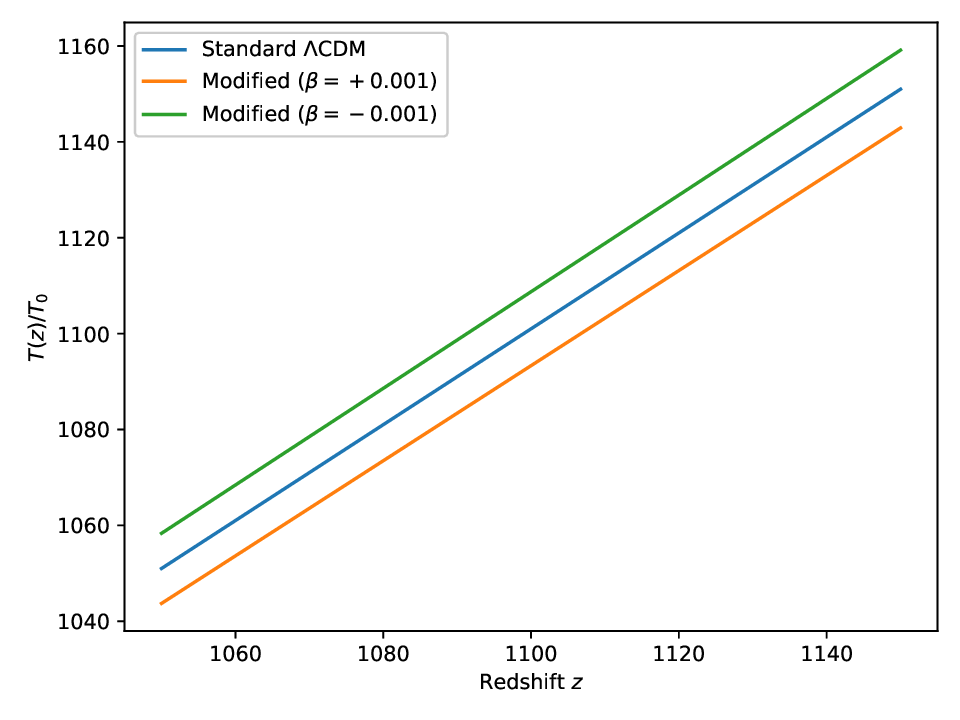}
\caption{CMB temperature–redshift relation in the recombination era. The solid curve shows the standard adiabatic scaling $T(z)=T_0 (1+z)$ while the dashed and dotted curves correspond to the modified evolution induced by EDE–photon coupling with $\beta=+0.001$ and $\beta=-0.001$, respectively. Although the deviation from the standard relation is small, it is systematic and becomes relevant at the epoch of last scattering. }
 \end{center}
\end{figure}
 with $\beta=\epsilon/4$. Taking $\epsilon$ as a constant corresponds to a regime in which $\phi\propto\ln a$ or $\epsilon\propto\dot{\phi}/H$. Such logarithmic evolution for $\phi$ is a characteristic feature of scaling or attractor solutions which naturally arise in cosmological models with exponential couplings or potentials and are common in interacting dark energy scenarios \cite{bis3}. In these attractor regimes, the scalar field dynamics becomes insensitive to initial conditions and maintains fixed ratios with the dominant background component over extended cosmological epochs. Explicit realizations of this behavior in interacting scalar field cosmologies including EDE models have been discussed in detail in the literature \cite{galli}. Consequently, the assumption of constant $\epsilon$ should be interpreted as an effective description of a pre-recombination attractor solution rather than a purely phenomenological ansatz.\\
From (\ref{a5}) and (\ref{a7}), one gets
 \begin{equation}
 T(z)=T_0 (1+z)^{1-\beta}
\label{a8}\end{equation}
Several authors have proposed modifications to the temperature–redshift relation of the form (\ref{a8}) arising from various cosmological mechanisms such as photon–axion conversion \cite{avg}, decaying vacuum models \cite{lima2} \cite{luzzi} or interactions within the dark sector \cite{hur}. In the present work, this modification, encapsulated in the parameter $\beta$, is attributed to the energy transfer between EDE component and the radiation bath. Apart from its technical simplicity, taking $\epsilon$ as a constant parameter has a further important advantage in connecting the present framework to the standard phenomenological description of non-adiabatic photon evolution given by (12). It is precisely the form commonly adopted in the literature to parametrize deviations from adiabatic photon dilution. In this sense, the assumption of constant $\epsilon$ provides a minimal and physically transparent parametrization that allows a direct comparison between the underlying EDE–photon interaction and existing observational constraints on modified CMB temperature evolution.

~~~~~~~~~~~~~~~~~~~~~~~~~~~~~~~~~~~~~~~~~~~~~~~~~~~~~~~~~~~~~~~~~~~~~~~~~~~~~~~~~~~~~~~~~~~~~~~~~~~~~~~~~~~~~~~~~~~~~~~~~~~~~~~~~~~~~~~~~~~~~
\section{Linear perturbations in the tight-coupling regime}
In this section we investigate the implications of the modified CMB temperature-redshift relation for linear scalar perturbations around recombination. The cosmic fluid in the present model consists of radiation which is dominant and subdominant components which consist of EDE characterizing by the scalar field $\phi$, tightly coupled baryons and decoupled CDM.
Although the EDE interacts with radiation and is responsible for modifying the background temperature-redshift relation, this interaction corresponds to a homogeneous energy exchange and does not induce momentum transfer or anisotropic stress at linear order. As a result, the EDE-photon coupling affects perturbations indirectly through the modified background evolution, recombination history and sound speed while the structure of the linear perturbation equations remains unchanged. In contrast, baryon-photon coupling involves direct momentum exchange through Thomson scattering and therefore enters explicitly at the perturbation level. This distinction allows us to consistently neglect scalar field perturbations at leading order to focus on the dynamics of the tightly coupled photon-baryon fluid.
~~~~~~~~~~~~~~~~~~~~~~~~~~~~~~~~~~~~~~~~~~~~~~~~~~~~~~~~~~~~~~~~~~~~~~~~~~~~~~~~~~~~~~~~~~~~~~~~~~~~~~~~~~~~~~~
\subsection{linear perturbations in Newtonian gauge}
We consider scalar perturbations of a spatially flat FRW spacetime in conformal Newtonian gauge
\begin{equation}
ds^2 = a^2(\eta)\left[-(1+2\Psi)d\eta^2 + (1-2\Phi)d\mathbf{x}^2\right]
\label{eq:metric}
\end{equation}
where $\Phi$ and $\Psi$ are gauge invariant gravitational potentials and $\eta$ denotes conformal time. At early times, anisotropic stress is negligible and we therefore set $\Phi=\Psi$.
For each component $i$ we define the density contrast $\delta_i=\delta\rho_i/\rho_i$ and the velocity divergence $ \theta_i =\nabla \cdot \mathbf{v}_i$. CDM contributes to the metric perturbations through the Einstein equations. Since it is pressureless and decoupled from radiation, it does not participate directly in acoustic oscillations. The scalar field $\phi$ whose background energy density remains small compared to radiation during the relevant epoch is treated analogously.
~~~~~~~~~~~~~~~~~~~~~~~~~~~~~~~~~~~~~~~~~~~~~~~~~~~~~~~~~~~~~~~~~~~~~~~~~~~~~~~~~~~~~~~~~~~~~~~~~~~~~~~~~~~~~~~~~~~~~~~~
\subsection{tight-coupling approximation}
Prior to recombination, frequent Thomson scattering tightly couples photons and baryons. For the Fourier modes relevant to acoustic oscillations, the interaction rate satisfies \cite{tight1}
\begin{equation}
a n_e \sigma_T \gg k
\label{eq:tight}
\end{equation}
where $n_e$ is the free electron density and $\sigma_T$ is the Thomson cross section. In this limit, photon and baryon velocities are locked
\begin{equation}
\theta_\gamma \simeq \theta_b \equiv \theta
\label{eq:vel_lock}
\end{equation}
and photon anisotropic stress is strongly suppressed. The photon--baryon system may therefore be described as a single effective fluid without introducing higher angular moments of the photon distribution.
~~~~~~~~~~~~~~~~~~~~~~~~~~~~~~~~~~~~~~~~~~~~~~~~~~~~~~~~~~~~~~~~~~~~~~~~~~~~~~~~~~~~~~~~~~~~~~~~~~~~~~~~~~~~~~~
\subsection{photon-baryon fluid equations}
Under the tight-coupling approximation, the linearized continuity equations for photons and baryons are \cite{tight2}
\begin{equation}
\delta_\gamma' = -\frac{4}{3}\theta + 4\Psi'
\label{eq:cont_g}
\end{equation}
\begin{equation}
\delta_b' = -\theta + 3\Psi'
\label{eq:cont_b}
\end{equation}
where primes denote derivatives with respect to conformal time. In addition to the continuity equations, the evolution of velocity perturbations in the photon and baryon components is governed by their respective Euler equations. These equations describe momentum conservation in the presence of gravitational potentials and include the effects of Thomson scattering which mediates momentum exchange between photons and baryons prior to recombination. In Fourier space and conformal time, they take the following standard forms
\begin{equation}
\theta_b' + \mathcal{H}\,\theta_b
=
k^2 \Psi
+
\frac{1}{R}\, a n_e \sigma_T
\left(\theta_\gamma - \theta_b\right)
\label{baryon_euler}
\end{equation}
\begin{equation}
\theta_\gamma'
=
k^2\left(\frac{1}{4}\delta_\gamma + \Psi\right)
-
a n_e \sigma_T
\left(\theta_\gamma - \theta_b\right)
\label{photon_euler}
\end{equation}
where $R\equiv \frac{3\rho_b}{4\rho_\gamma}$ and $\mathcal{H}=a'/a$.
Combining the latter equations in the limit characterized by (\ref{eq:vel_lock}),
yields a single momentum conservation equation for the photon–baryon fluid
\begin{equation}
\theta' + \mathcal{H}(1+R)\theta
= k^2\left(\frac{\delta_\gamma}{4(1+R)} + \Psi\right)
\label{eq:euler}
\end{equation}
The modified temperature-redshift relation (\ref{a8}) implies a modified scaling of the photon energy density
\begin{equation}
\rho_\gamma(z)\propto (1+z)^{4(1-\beta)}
\label{eq:rhog}
\end{equation}
Consequently, the baryon loading parameter $R(z)$ and the sound speed of the photon--baryon fluid acquire a modified redshift dependence.
Combining equations (\ref{eq:cont_g})-(\ref{eq:euler}), one obtains the acoustic oscillator equation
\begin{equation}
\delta_\gamma'' + \frac{R'}{1+R} \delta_\gamma' + c_s^2 k^2 \delta_\gamma = -4\left(\Psi'' + \frac{R'}{1+R}\Psi'\right)
\label{eq:osc}
\end{equation}
where the sound speed is
\begin{equation}
c_s^2 = \frac{1}{3(1+R)}
\label{eq:cs}
\end{equation}
The equation (23) summarizes the dynamics of scalar perturbations in the tightly coupled photon–baryon fluid prior to recombination. It shows explicitly that density perturbations undergo acoustic oscillations whose frequency is governed by the sound speed and whose phase is determined by the sound horizon at decoupling. Since both quantities depend on the background radiation density, any modification to the temperature–redshift relation induced by EDE–photon coupling feeds directly into the evolution of these oscillations. As a result, (23) provides the key link between the modified thermal history discussed above and the observable imprints on the CMB anisotropy spectrum. In the following, we exploit this connection to identify the primary physical observables (such as the sound speed, sound horizon and angular acoustic scale) and to assess how they are altered in the present model and constrained by CMB observations.
~~~~~~~~~~~~~~~~~~~~~~~~~~~~~~~~~~~~~~~~~~~~~~~~~~~~~~~~~~~~~~~~~~~~~~~~~~~~~~~~~~~~~~~~~~~~~~~~~~~
\subsection{physical observables and acoustic scales}
A central quantity controlling the phase of acoustic oscillations is the comoving sound horizon at recombination defined as
\begin{equation}
r_s(z_{\rm rec}) = \int_0^{\eta_{\rm rec}} c_s(\eta)\, d\eta
\label{eq:rs}
\end{equation}
where $c_s$ is the sound speed of the tightly coupled photon–baryon fluid and $\eta_{rec}$ denotes the conformal time at recombination. Physically, $r_s$ represents the maximum distance a sound wave can travel in the primordial plasma before photons decouple from baryons. The acoustic oscillator equation directly implies that the phase of density perturbations at recombination is proportional to $k r_s$ establishing the sound horizon as the fundamental physical scale that determines the spacing of acoustic features in the CMB temperature and polarization spectra.\\
The EDE–photon coupling alters the background evolution of the radiation energy density by modifying the temperature–redshift relation. This modification affects both the sound speed and the expansion rate entering the sound-horizon integral in (24). This leads to a shift in the sound horizon relative to its standard $\Lambda$CDM value summarized in the equation (25) below. For small deviations from the standard temperature scaling (characterized by the parameter $\beta$) the resulting change in the sound horizon can be expressed schematically as
\begin{equation}
\frac{\Delta r_s}{r_s} \sim \mathcal{O}(\beta)
\label{eq:drs}
\end{equation}
indicating that the effect is perturbative but cumulative over the pre-recombination epoch. It reflects the fact that even a small modification in the thermal history can integrate into a non-negligible shift in the acoustic scale \cite{Hu1}.\\
The scaling in (25) comes from how both the sound speed $c_s$ and the expansion rate $H(z)$ acquire small and linear corrections linear in $\beta$ through the modified radiation density $\rho_{\gamma}\propto (1+z)^{4(1-\beta)}$. Expanding the integrand of (24) to first order in $\beta$ reveals that the integrand is perturbed by $\delta(c_s/H)/(c_s/H)\sim \mathcal{O}(\beta)$. Since the sound horizon accumulates over a wide redshift interval in radiation-dominated era, these small corrections integrate coherently that lead to a net fractional shift $\Delta r_s/r_s\sim \mathcal{O}(\beta)$.\\
The observational relevance of the sound horizon becomes manifest when considering the form of photon density perturbations at recombination. To leading order, solutions of the acoustic oscillator equation take the form \cite{tight2} \cite{Hu1}
\begin{equation}
\delta_\gamma(k,\eta_{\rm rec}) \sim \cos\!\left[k r_s(z_{\rm rec})\right]
\label{eq:phase}
\end{equation}
showing explicitly that the phase of the oscillations (and hence the positions of the acoustic peaks in multipole space) is determined by the sound horizon. High-precision CMB measurements, most notably from the Planck satellite, tightly constrain this phase and therefore place stringent bounds on any mechanism that alters $r_s$ \cite{planck}.
Equations (25)–(26) thus provide the direct link between the perturbation dynamics discussed in 3.3 and the observable CMB anisotropy spectrum. Here they clarify how a modified temperature–redshift relation feeds into the acoustic scale without introducing new dynamical degrees of freedom at the perturbation level. In the following, we use this connection to discuss how current CMB observations constrain the allowed magnitude of the EDE–photon coupling parameter $\beta$.

~~~~~~~~~~~~~~~~~~~~~~~~~~~~~~~~~~~~~~~~~~~~~~~~~~~~~~~~~~~~~~~~~~~~~~~~~~~~~~~~~~~~~~~~~~~~~~~~~~~~~~~~~~~~~~~~~~~~~~~~~~~~~~~~
\subsection{Sound speed, sound horizon and implications for the Hubble tension}
In the standard $\Lambda$CDM model, radiation redshifts as $\rho_\gamma \propto
a^{-4}$ while baryons scale as $\rho_b \propto a^{-3}$ implying $R \propto a$.
In the present model, however, EDE-photon coupling modifies the radiation
dilution law according to
\begin{equation}
\rho_\gamma \propto a^{-4+\epsilon}
\label{eq:rhogamma_mod}
\end{equation}
which leads to a modified evolution of the baryon loading parameter
\begin{equation}
R(a) \propto a^{1-\epsilon}
\label{eq:Rmod}
\end{equation}
Therefore, the redshift dependence of the sound speed is altered through the modified background
evolution of $\rho_\gamma$. The sign of the coupling parameter $\beta=\epsilon/4$
determines the direction of this effect: for $\beta>0$, radiation redshifts more
slowly suppressing the growth of $R$ and mildly enhancing the sound speed. In contrast, for $\beta<0$ radiation dilutes more rapidly increasing $R$ and reducing $c_s$.\\
The cumulative impact of this modified sound speed is encoded in the comoving sound
horizon at recombination defined as
\begin{equation}
r_s(z_\ast) = \int_{z_\ast}^{\infty} \frac{c_s(z)}{H(z)}\,dz
\label{eq:rs_def}
\end{equation}
where $z_\ast$ denotes the redshift of recombination. Because both the sound speed and the expansion rate inherit a
dependence on $\beta$ through the modified radiation density, the sound horizon is
shifted relative to its standard $\Lambda$CDM value.
The observational relevance of the sound horizon arises from its role in setting the
angular acoustic scale
\begin{equation}
\theta_s = \frac{r_s(z_\ast)}{D_A(z_\ast)}
\label{eq:theta_star}
\end{equation}
which determines the spacing of the acoustic peaks in the CMB temperature and
polarization spectra. Planck observations constrain $\theta_\ast$ at the
$10^{-3}$ level which makes it one of the most precisely measured quantities in
cosmology. In the present framework, the dominant effect of the EDE-photon
coupling enters through the modification of $r_s$ while changes to the angular
diameter distance $D_A$ are subleading at early times. \emph{As a result, positive values
of $\beta$ which reduce the sound horizon, tend to shift the CMB-inferred value of
the Hubble constant upward and partially alleviates the Hubble tension.}
\begin{figure}[t]
\centering
\includegraphics[width=0.8\linewidth]{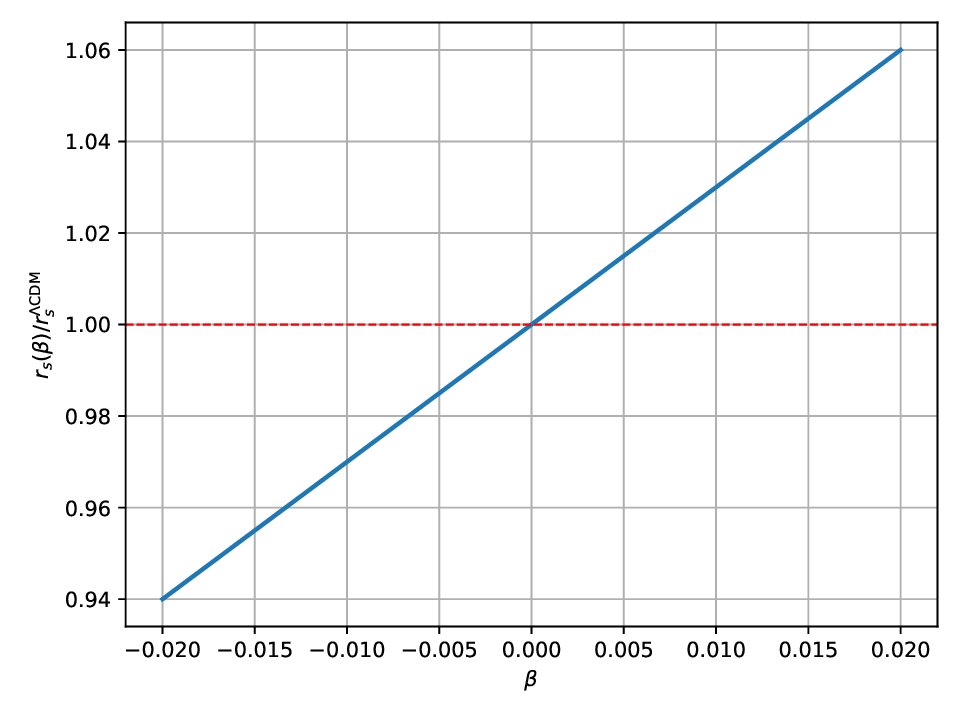}
\caption{Ratio of the comoving sound horizon at recombination in the presence of EDE--photon coupling $r_s(\beta)$ to its standard $\Lambda$CDM value as a function of the coupling parameter $\beta$. The result is shown in the small-coupling regime $\beta\ll 1$ where the fractional change in the sound horizon scales linearly with $\beta$. The horizontal dashed line indicates the standard $\Lambda$CDM prediction.}
\label{fig:rs_beta}
\end{figure}

~~~~~~~~~~~~~~~~~~~~~~~~~~~~~~~~~~~~~~~~~~~~~~~~~~~~~~~~~~~~~~~~~~~~~~~~~~~~~~~~~~~~~~~~~~~~~~~~~~~~~~~~

\subsection{comparison with Planck observations}
We have shown that the EDE-photon coupling modifies the background evolution of the radiation sector and leads to a non-standard temperature-redshift relation characterized by the parameter $\beta$. This modification affects the photon energy density and consequently alters the sound speed of the tightly coupled photon-baryon fluid prior to recombination. Through its cumulative effect during the radiation-dominated era, the coupling therefore modifies the comoving sound horizon at last scattering.
The sound horizon at recombination is defined as
\begin{equation}
r_s(\beta)=\int_{z_*}^{\infty}\frac{c_s(z;\beta)}{H(z;\beta)}\,dz ,
\end{equation}
where $z_*$ denotes the redshift of recombination and $H(z)$ is the Hubble expansion rate. Both quantities inherit a dependence on $\beta$ through the modified temperature-redshift relation and the resulting change in the radiation energy density. The form of the sound horizon integral remains unchanged and only the background quantities entering the integrand are modified.
For small values of the coupling parameter, $|\beta|\ll 1$, the effect on the sound horizon can be treated perturbatively. Expanding around the standard $\Lambda$CDM result yields
\begin{equation}
\frac{r_s(\beta)}{r_s^{\Lambda{\rm CDM}}}
\simeq
1+\mathcal{F}\,\beta ,
\end{equation}
where $\mathcal{F}$ is a dimensionless coefficient of order unity that encodes the integrated sensitivity of the sound horizon to changes in the radiation background. This expression makes explicit that the fractional shift in the sound horizon is linear in $\beta$ for sufficiently small coupling.
The linear dependence of the sound horizon on the coupling parameter is illustrated in Fig. 2 where the ratio $r_s(\beta)/r_s^{\Lambda{\rm CDM}}$ is plotted as a function of $\beta$. The figure highlights both the small magnitude of the effect and the importance of the sign of $\beta$ with positive (negative) values corresponding to an increase (decrease) of the sound horizon relative to its standard value.\\
Here the dominant effect of the EDE-photon coupling enters through the modification of $r_s$ while changes to $D_A$ are subleading at early times.
Requiring the predicted angular acoustic scale to remain consistent with the Planck measurement therefore implies
\begin{equation}
\left|\frac{\Delta r_s}{r_s}\right|\lesssim 10^{-3}
\end{equation}
which using the linearized expression above translates into an upper bound on the coupling parameter
\begin{equation}
|\beta|\lesssim\frac{10^{-3}}{|\mathcal{F}|}
\end{equation}
Given that $\mathcal{F}\sim\mathcal{O}(1)$, this result places a stringent constraint on the magnitude of the EDE-photon coupling, $|\beta|\lesssim 10^{-3}$.
This bound should be interpreted as a semi-analytic consistency constraint derived from the precision with which Planck observations fix the acoustic scale. A full quantitative determination of the allowed parameter space would require implementing the modified background evolution in a Boltzmann solver and performing a likelihood analysis which lies beyond the scope of the present work. Nevertheless, the analysis demonstrates that even small deviations from the standard temperature-redshift relation can produce observable effects in early Universe probes.
~~~~~~~~~~~~~~~~~~~~~~~~~~~~~~~~~~~~~~~~~~~~~~~~~~~~~~~~~~~~~~~~~~~~~~~~~~~~~~~~~~~~~~~~~~~~~~~~~~~~~~~~~~~~~~~~~~~~~~~~~~~~~~~~~~~~~
\section{Conclusions}
In this work we have studied the cosmological implications of an EDE scalar field interacting with the radiation sector prior to recombination with particular emphasis on its impact on the CMB temperature–redshift relation and acoustic-scale physics. The motivation for this analysis is rooted in the growing body of evidence that the standard $\Lambda$CDM model, despite its remarkable success, faces persistent internal inconsistencies. \\
First, we focused on the background evolution and the thermal history of the Universe. Starting from an explicit EDE–photon coupling in the action, we derived a modified radiation dilution law and showed that it leads naturally to a non-standard CMB temperature–redshift relation of the form $T(z)=T_0 (1+z)^{1-\beta}$. This modification admits a clear physical interpretation as homogeneous photon creation or annihilation depending on the sign of the coupling parameter $\beta$ and connects our model to earlier phenomenological studies of non-adiabatic photon evolution [10]–[14]. Importantly in contrast to purely phenomenological parameterizations, the deviation here arises from a well-defined early Universe interaction motivated by scalar field dynamics.\\
Second, we investigated linear scalar perturbations in the tight-coupling regime relevant for CMB acoustic physics. A central result of this analysis is that the EDE–photon interaction does not introduce new propagating degrees of freedom at the perturbation level. Instead, its impact enters entirely through modifications to the background quantities such as the radiation energy density, the baryon loading parameter, the sound speed and ultimately the comoving sound horizon at recombination. The structure of the photon–baryon perturbation equations therefore remains unchanged consistent with previous analyses of interacting but momentum-conserving dark sectors [15] [16] while the acoustic phase acquires a calculable dependence on the coupling strength.\\
The dominant observational consequence of this framework is a shift in the sound horizon which directly affects the angular acoustic scale measured with high precision by Planck [3]. Requiring consistency with the observed acoustic peak positions leads to a semi-analytic bound $\beta\lesssim 10^{-3}$ indicating that even percent-level deviations from adiabatic photon dilution are strongly constrained. This result reinforces a key lesson of modern precision cosmology: the pre-recombination thermal history is exquisitely sensitive to new physics and small background modifications can leave detectable imprints in early Universe observables.
From the perspective of the Hubble tension, these findings are particularly relevant. A reduction of the sound horizon prior to recombination is a well-known mechanism for reconciling the CMB-inferred value of $H_0$ with late-time distance ladder measurements [7]. In the present model, positive values of $\beta$ naturally lead to a smaller sound horizon without altering the perturbation structure thereby shifting the inferred value of $H_0$ upward in a controlled manner. This places the EDE–photon interaction studied here in the same conceptual class as other early time solutions to the Hubble tension including EDE models [7], additional relativistic species and modified pre-recombination expansion histories while offering a distinct physical origin rooted in interactions of the radiation sector.\\
Beyond the Hubble tension, the results also have potential implications for other cosmological discrepancies, most notably the $S_8$ tension between CMB-inferred and low-redshift measurements of structure growth as highlighted by weak-lensing surveys such as KiDS, DES, HSC and their comparison with Planck data [7] \cite{kid}.
 Although our analysis did not explicitly address late time clustering, modifications to the early radiation background and expansion rate can indirectly affect the calibration of matter fluctuations through changes in the sound horizon and the mapping between early and late time observables. Similar mechanisms have been shown to partially alleviate tensions in $S_8$ in models where early time physics alters the normalization of matter perturbations without invoking strong late-time modifications [7[ [15]. A full assessment of this possibility within the present framework would require extending the analysis beyond recombination and including large-scale structure data.\\
The relevance of these results extends naturally to future observational programmes. Upcoming CMB experiments with improved sensitivity to acoustic scales, polarization and damping-tail physics will further tighten constraints on deviations from the standard temperature–redshift relation. Independent probes of the CMB temperature evolution at intermediate redshifts such as Sunyaev–Zel’dovich measurements [18] [19] offer complementary tests of the model and can help disentangle early time interactions from late time opacity or exotic photon-loss mechanisms. Together, these observations will sharpen our ability to test interacting EDE scenarios as viable resolutions to current cosmological tensions.\\\\
~~~~~~~~~~~~~~~~~~~~~~~~~~~~~~~~~~~~~~~~~~~~~~~~~~~~~~~~~~~~~~~~~~~~~~~~~~~~~~~~~~~~~~~~~~~~~~~~~~~~~~~~~~~~~~~~~~~~~~~~~

\end{document}